\documentclass[proceedings,nohyper]{JHEP} % 10pt is ignored!

\usepackage{epsfig}                   % please use epsfig for figures

\newcommand{\cB}{{\cal B}}

\newcommand{\hf}{\frac12}

\newcommand{\bea}{\begin{eqnarray}}
\newcommand{\eea}{\end{eqnarray}}
\newcommand{\be}{\begin{equation}}
\newcommand{\ee}{\end{equation}}
\newcommand{\bt}{\begin{tabular}}
\newcommand{\et}{\end{tabular}}
\newcommand{\ba}{\begin{array}}
\newcommand{\ea}{\end{array}}

\newcommand{\Tr}{\mathop{\rm Tr}}
\newcommand{\Pf}{\mathop{\rm Pf}}

\newcommand{\lra}{\leftrightarrow}
\def \msmall#1{{\scriptstyle{#1}}}

\def \half{{\textstyle\hf}}

\def \soll={\stackrel{!}{=}}
\def \rf=#1{\stackrel{(\ref{#1})}{=}}

\def \Wtree{W_{\rm tree}}

\def \Wnp{W_{\rm np}}
\def \mag{_{\rm mag}}
\def \Wmag{W\mag}

\def \w{{\rm with\ }}

\def \barfill{\leaders\hrule height 0.1 true pt\hfill}
\def \overbar#1{\vbox{\ialign{##\crcr\barfill\crcr\noalign{\kern 1pt
                                      \nointerlineskip}$\hfil{#1}\hfil$\crcr}}}
\def \scriptbar#1{{\vbox{\ialign{##\crcr\thinspace\barfill\thinspace\crcr
    \noalign{\kern 0.8pt\nointerlineskip}$\hfil{\scriptstyle #1}\hfil$\crcr}}}}

\newlength{\quadlength} 
\newlength{\abstand} \newlength{\breite}

\def \larrow#1{\count1=0\leftarrow\loop\ifnum\count1<#1\hspace{-0.1em}
               \joinrel\relbar\advance\count1 by 1\repeat}
\def \rarrow#1{\count1=0\loop\ifnum\count1<#1\hspace{-0.1em}
               \relbar\joinrel\advance\count1 by 1\repeat\rightarrow}
\def \lrarrow#1{\count1=0\leftarrow\loop\ifnum\count1<#1\hspace{-0.1em}
                \joinrel\relbar\advance\count1 by 1\repeat\joinrel\rightarrow}

\newcommand{\drawsquare}[2]{\hbox{%
\rule{#2pt}{#1pt}\hskip-#2pt%  left vertical
\rule{#1pt}{#2pt}\hskip-#1pt%  lower horizontal
\rule[#1pt]{#1pt}{#2pt}}\rule[#1pt]{#2pt}{#2pt}\hskip-#2pt%  upper horizontal
\rule{#2pt}{#1pt}}% right vertical

\newcommand{\Ysym}{\raisebox{-.5pt}{\drawsquare{6.5}{0.4}}\hskip-0.4pt%
        \raisebox{-.5pt}{\drawsquare{6.5}{0.4}}}%  symmetric second rank
\newcommand{\Yasym}{\raisebox{-3.5pt}{\drawsquare{6.5}{0.4}}\hskip-6.9pt%
        \raisebox{3pt}{\drawsquare{6.5}{0.4}}}%  antisymmetric second rank

\def \Ysymb{\overline{\Ysym}}

\def \Yasymb{\overline{\Yasym}}

\def \nonpert{non\discretionary{-}{}{-}per\-tur\-ba\-tive\ }

\def \nonab{non\discretionary{-}{}{-}Abel\-ian\ }
\def \nonvan{non\discretionary{-}{}{-}van\-ish\-ing\ }
\def \nonren{non\discretionary{-}{}{-}re\-nor\-ma\-li\-za\-tion\ }
\def \elmag{e\-lec\-tric\discretionary{-}{}{-}mag\-ne\-tic\ }
\def \afun{antifundamental\ }
\def \asym{antisymmetric\ }

\conference{Trieste Meeting of the TMR Network on Physics beyond the SM}

\title{Confining N=1 SUSY gauge theories from Seiberg duality}

\author{Matthias Klein\\
        Departamento de F\'\i sica Te\'orica C-XI and 
        Instituto de F\'\i sica Te\'orica C-XVI \\
        Universidad Aut\'onoma de Madrid, Cantoblanco, 28049 Madrid, Spain\\
        E-mail: \email{Matthias.Klein@uam.es}}

\abstract{In this talk I review and generalize an idea of Seiberg that an 
   $N=1$ supersymmetric gauge theory shows confinement without breaking of 
   chiral symmetry when the gauge symmetry of its magnetic dual is completely 
   broken by the Higgs effect. It is shown how the confining spectrum of 
   a supersymmetric gauge theory can easily be derived when a magnetic dual
   is known and this method is applied to many models containing fields in 
   second rank tensor representations and an appropriate tree-level 
   superpotential.}

\begin{document} 

\section{Introduction}
Due to holomorphicity properties and \nonren theorems valid in
$N=1$ supersymmetric theories it has become possible to argue that some 
supersymmetric gauge theories with special matter content confine at low 
energies. The first example is due to Seiberg \cite{Seib_D49} who found 
that supersymmetric quantum chromodynamics (SQCD) with gauge group $SU(N_c)$ 
and $N_f$ quark flavors shows confinement when $N_f=N_c$ or $N_f=N_c+1$. 
This has been generalized to more complicated models. All $N=1$ 
supersymmetric gauge theories with vanishing tree-level superpotential 
which confine at low energies could be classified \cite{confine,qmm,
affineqm} because they are constrained by an index argument.
When a tree-level superpotential is present the index argument is no longer
valid. Because of the lower symmetry the non-perturbative superpotential is
less constrained and one expects more confining models to exist. Indeed,
Cs\'aki and Murayama \cite{newconf} showed that many of the Kutasov-like 
\cite{KuSch} models exhibit confinement for special values of the number 
of quark flavors $N_f$.\footnote{Some further confining models with \nonvan 
tree-level superpotential are discussed in \cite{TaYo}.} 
These models contain fields in tensor representations of the gauge group
and an appropriate superpotential for these tensor fields. For all of 
the models considered in \cite{newconf} a dual description in terms of
magnetic variables is known \cite{KuSch,Intri,LS,ILSdual} and the authors of 
\cite{newconf} used the fact that the electric gauge theory confines when
its magnetic dual is completely higgsed.

Seiberg already used this idea as an additional consistency check in his 
original paper establishing \elmag duality for \nonab 
$N=1$ supersymmetric gauge theories \cite{Seib}. He showed how the confining 
superpotential of SQCD with $N_f=N_c+1$ could be obtained by a perturbative 
calculation in the completely broken magnetic gauge theory. Under duality the 
fields of the magnetic theory (which are gauge singlets as the gauge 
symmetry is completely broken) are mapped to the mesons and baryons of 
the electric theory and the confining superpotential is easily shown 
to be the image of the magnetic superpotential under this mapping \cite{Seib}.
This a realization in $N=1$ supersymmetric gauge theories of an old idea of
't Hooft and Mandelstam \cite{dualmeiss} that confinement is driven by 
condensation of magnetic monopoles.

Now, many gauge theory models have been found that possess a dual description
in terms of magnetic variables in the infrared. This allows us to predict 
many new examples of confining gauge theories. The idea described in the
previous paragraph was first used by the authors of \cite{KSS} to determine 
the confining spectrum of the model proposed by Kutasov \cite{KuSch} and has 
been applied by Cs\'aki and Murayama \cite{newconf} to six further 
models that confine in the presence of an appropriate superpotential.

In this talk I review the original example of Seiberg \cite{Seib} and
explain how \elmag duality is used to obtain the low-energy 
spectrum and the form of the \nonpert superpotential of confining gauge 
theories \cite{mklein}. One finds that all of the gauge theory models based 
on simple gauge groups considered in \cite{ILSdual,patterns} confine when the 
gauge groups of their magnetic duals are completely broken by the Higgs 
effect. For nine of these theories the confining phase has not been discussed
before.

\section{Phase structure of SQCD}
Let us briefly review the well-known phase structure of SQCD \cite{ISlectures}.
By this we mean an $N=1$ supersymmetric $SU(N_c)$ gauge theory with $N_f$ 
quark flavors, i.e.\ $N_f$ chiral matter supermultiplets $Q$ transforming in 
the fundamental representation of the gauge group and the same amount of matter
multiplets $\bar Q$ transforming in the \afun representation. 
Consider first the case of vanishing tree-level superpotential. Depending on 
the relative values of $N_f$ and $N_c$ the the low-energy theory resides in 
different phases.

$N_f=0$: This is pure super Yang-Mills theory. It is believed to show
confinement. According to an index argument by Witten \cite{W-index} there
are $N_c$ distinct supersymmetric vacua.

$0<N_f<N_c$: There is a \nonpert superpotential generated by gluino
condensation (for $N_f<N_c-1$) or by instantons (for $N_f=N_c-1$),
as was shown by Affleck, Dine and Seiberg \cite{ADS}:
\be \label{WADS}
\Wnp=(N_c-N_f)\left(\Lambda^{3N_c-N_f}\over\det M\right)^{1\over N_c-N_f}\,,
\ee
where $\Lambda$ is the dynamically generated scale of the theory and the
meson matrix $M$ is defined by $M^{ij}=Q^{\alpha i}\bar Q^j_\alpha$,
$i,j=1,\ldots,N_f$, $\alpha=1,\ldots,N_c$. The minimum of the potential lies
at infinite field expectation values and therefore the theory has no stable
vacuum for this range of parameters.

$N_f=N_c$: In this case the superpotential vanishes even at the \nonpert level
\cite{Seib_D49}. As a consequence the flat directions that pa\-ram\-e\-trize 
the moduli space of vacua are not lifted in the quantum theory. The low-energy
spectrum is given by the mesons $M^{ij}$ defined above and the baryons
$B=\det Q$, $\bar B=\det\bar Q$ (the quarks $Q$, $\bar Q$ are viewed as
$(N_f\times N_c)$-matrices). The physical degrees of freedom at low energies 
being gauge invariant means that the theory confines. The classical
constraint $\det M=B\bar B$ is modified in the quantum theory \cite{Seib_D49} 
to
\be \label{qconstr}
\det M - B\bar B = \Lambda^{2N_c}\,.
\ee
The expectation values of the mesons and baryons that satisfy this constraint 
span the quantum moduli space. The observation that the expectation values
of $\det M$ and $B\bar B$ cannot vanish simultaneously tells us that the
chiral symmetry is spontaneously broken.

$N_f=N_c+1$: There is again a quantum moduli space, but now the classical
constraints are not modified by quantum effects. In the low-energy theory
they can be derived from the \nonpert superpotential \cite{Seib_D49}
\be \label{Wconf}
\Wnp={\bar BMB-\det M\over\Lambda^{2N_f-3}}\,,
\ee
where the baryons $B^i$ are defined as the determinant of the quark matrix $Q$
with the $i$-th line omitted. This describes confinement without \linebreak
breaking of the chiral symmetry.

$N_c+2\le N_f\le {3\over2}N_c$: The low-energy energy theory is rather
complicated and more appropriately described in terms of dual magnetic
variables. The dual magnetic theory is infrared free for this range of
parameters.

${3\over2}N_c<N_f<3N_c$: At low energies the theory is driven to an infrared
fixed point of the renormalization group \cite{Seib} and resides in a \nonab
Coulomb phase. Seiberg found a dual description of this model \cite{Seib} by
an $SU(N_f-N_c)$ gauge theory with $N_f$ (magnetic) quark flavors $q$, $\bar q$
and $N_f^2$ additional singlets $M\mag^{ij}$ which couple to the magnetic
quarks via the superpotential
\be \label{Wmag}   \Wmag=M\mag q\bar q\,.
\ee
This magnetic theory flows to the same infrared fixed point. The gauge
invariant operators of both theories are in one-to-one correspondence:
\bea \label{dualmap}
M &\ \lrarrow{1}\ &\mu\,M\mag\,,\\
B &\ \lrarrow{1}\ &\sqrt{-(-\mu)^{N_c-N_f}\Lambda^{3N_c-N_f}}\:B\mag\,.
\nonumber
\eea
The mass scale $\mu$ had to be introduced by dimensional analysis. The three
scales $\Lambda$, $\Lambda\mag$ and $\mu$ are related by
\be \label{scales}
\Lambda^{3N_c-N_f}\Lambda\mag^{3(N_f-N_c)-N_f}=(-1)^{N_f-N_c}\mu^{N_f}\,.
\ee

$N_f>3N_c$: In the infrared the theory flows to the trivial fixed point of
free quarks and gluons.

\section{Confinement from duality}
\FIGURE{$
\begin{array}{ccc}
SU(N_c),N_f\ {\rm flavors}   &\quad\stackrel{\rm duality}{\lrarrow{6}}\quad 
                             &SU(N_f-N_c),N_f\ {\rm flavors}\\
W=\Tr(mM)                    &&\Wmag=M\mag q\bar q+\mu\Tr(mM\mag)\\
\Bigg\downarrow              &&\Bigg\downarrow\\
SU(N_c),N_f-p\ {\rm flavors} &\quad\stackrel{\rm duality}{\lrarrow{6}}\quad
                             &SU(N_f-N_c-p),N_f-p\ {\rm flavors}\\
\hat W=0                     &&\hat\Wmag=\hat M\mag \hat q\hat{\bar q}
\end{array}
\caption{The electric theory is deformed by adding mass terms for some of the
quarks and the magnetic theory is deformed correspondingly. After having
integrated out the massive modes one finds two effective theories that are
again dual to each other. Displayed are the tree-level contributions to the
superpotentials.}
$}
Let us consider deformations \cite{Seib} of the theory described in the 
previous section by mass terms $W=\Tr(mM)$, where $m$ is an $(N_f\times N_f)$-%
matrix of rank $p$. By the duality mapping (\ref{dualmap}) this corresponds to
adding a term $\mu\Tr(mM\mag)$ to the superpotential (\ref{Wmag}) in the 
magnetic theory (cf figure 1). As our treatment of SQCD is restricted to
the (Wilsonian) low-energy effective action we have to integrate out the 
massive modes from the deformed model. In the electric theory this just leads 
to a reduction of the number of quark flavors by $p$. Therefore the low-energy
theory is an $SU(N_c)$ gauge theory with $N_f-p$ quark flavors and vanishing 
superpotential $\hat W=0$. In the magnetic theory, integrating out the massive
components of $M\mag$ leads to \nonvan expectation values for $q$, $\bar q$ 
and thus the gauge symmetry is broken spontaneously. The low-energy theory is 
an $SU(N_f-N_c-p)$ gauge theory with $N_f-p$ quark flavors and superpotential
$\hat\Wmag=\hat M\mag\hat q\hat{\bar q}$, where hats denote the low-energy 
fields. One finds that the two effective theories are again dual to each other
\cite{Seib}, as shown in figure 1.

It is interesting to consider the special case $p=N_f-N_c-1$. Then one has an
effective $SU(N_c)$ gauge theory with $N_c+1$ quark flavors on the electric
side. The magnetic gauge symmetry is completely broken by the Higgs effect.
However, one color component of each of the $N_c+1$ quark flavors stays 
massless after the symmetry breaking. These $2(N_c+1)$ gauge singlets are
denoted by $\hat q$, $\hat{\bar q}$. They couple to the meson singlets 
$\hat M\mag$ via the tree-level superpotential (\ref{Wmag}). Due to
\nonren theorems this is not corrected in perturbation theory. But there are
\nonpert corrections generated by instantons. The full superpotential of the
low-energy magnetic theory reads \cite{Seib}
\be \label{Wmaglow}
\hat\Wmag=\hat M\mag\hat q\hat{\bar q}+\Lambda\mag^{3-N_f}\det\hat M\mag.
\ee
From the fact that all physical degrees of freedom of the effective magnetic 
theory are gauge invariant one expects that, as a consequence of the duality
mapping, the degrees of freedom of the electric theory are gauge singlets as 
well. We will see that this intuition is right. The mapping (\ref{dualmap})
gives
\bea \label{dualmaplow}
M &\ \lrarrow{1}\ &\mu\,M\mag\,,\\
B &\ \lrarrow{1}\ &\sqrt{\mu^{-1}\Lambda^{2N_f-3}}\:\hat q\,,\nonumber
\eea
and the scale matching (\ref{scales}) now reads
\be \label{scaleslow}
\Lambda^{2N_f-3}\Lambda\mag^{3-N_f}=-\mu^{N_f}\,.
\ee
The effective electric theory is described by the mesons $M$ and the baryons
$B$, $\bar B$. One can check that the 't Hooft anomaly matching conditions
\cite{tH} are satisfied for this low-energy spectrum. These conditions require
that if one gauges the global symmetries of the theory then the values of
the various triangle anomalies (which in general will not vanish) must 
coincide for the microscopic description in terms of quarks and gluons and
the macroscopic description in terms of mesons and baryons. Performing this
calculation one finds that the mesons and baryons obtained from the duality
mapping (\ref{dualmaplow}) are just the right degrees of freedom to match
the global anomalies of the microscopic theory. This means that the theory is
in the confining phase \cite{Seib_D49}, in agreement with the result for
$N_f=N_c+1$ of the previous section. It is easy to determine the full
superpotential of the effective electric theory by applying the duality
mapping (\ref{dualmaplow}) to the magnetic superpotential (\ref{Wmaglow}).
Using the scale relation (\ref{scaleslow}) one finds
\be \hat W={\bar BMB-\det M\over\Lambda^{2N_f-3}}\,,
\ee
which coincides with the superpotential (\ref{Wconf}) found in the previous 
section for $N_f=N_c+1$.

\section{Generalizations}
The results on SQCD described in the previous sections have been generalized
to other gauge theory models involving different gauge groups and/or
matter fields transforming in representations other than the fundamental.
For each of these models the electric theory shows confinement without 
breaking of the chiral symmetry when the gauge symmetry of its magnetic dual
is completely broken.

\subsection{other gauge groups}
The simplest extension of the results on SQCD described above consists in 
gauge theories with orthogonal or symplectic gauge groups. 
Let us first consider an $SO(N_c)$
gauge theory with $N_f$ matter fields $Q$ (quarks) transforming in the vector
representation of the gauge group and vanishing tree-level superpotential.
This has a dual description \cite{Seib,SOdual} in terms of a magnetic
$SO(N_f+4-N_c)$ gauge theory with $N_f$ quarks $q$ transforming in the
vector representation and $\half N_f(N_f+1)$ meson singlets $M\mag$ that 
couple to the magnetic quarks via $\Wmag=M\mag qq$. The gauge invariant
operators of both theories are in one-to-one correspondence to each other by a
mapping very similar to (\ref{dualmap}). For $N_f=N_c-3$ the magnetic theory
is completely higgsed, and one finds that the electric theory confines. The
confining spectrum (mesons $M$ and baryons $B$) as well as the correct 
confining superpotential 
\be W={MBB\over\Lambda^{2N_f+3}} \ee
can be obtained from the effective magnetic theory via the duality mapping. 

An $Sp(2N_c)$ gauge theory with $2N_f$ quarks $Q$ transforming in the 
fundamental representation and vanishing tree-level superpotential can 
equivalently be described \cite{IP} by an $Sp(2(N_f-2-N_c))$ gauge 
theory with $2N_f$ quarks $q$ and $N_f(2N_f-1)$ meson singlets $M\mag$ that 
couple to the magnetic quarks via $\Wmag=M\mag qq$. The duality mapping
between the gauge invariant operators of the two theories is simply given
by $M\:\lra\:\mu\,M\mag$; there are no baryons in symplectic gauge theories.
For $N_f=N_c+2$ the magnetic theory is completely higgsed, and one finds that 
the electric theory confines. The confining spectrum (mesons $M$) can be 
obtained from the effective magnetic theory via the duality mapping.
To obtain the correct confining superpotential 
\be W={\Pf M\over\Lambda^{2N_f-3}} \ee
more care is needed, because it is due to instanton corrections in the
effective magnetic gauge theory.

\subsection{gauge theories containing tensor fields}
For any $N=1$ supersymmetric model with vanishing tree-level superpotential
the form of the most general superpotential that can possibly be generated by
\nonpert effects is com\-plete\-ly fixed by the requirement that it be 
invariant under all symmetries of the considered model \cite{ADS,ILS,confine}.
For a theory with gauge group $G$ and chiral matter fields $\phi_l$ in 
representations $r_l$ of $G$ and dynamically generated scale $\Lambda$ 
one finds
\be \label{Weff}
W \propto \left({\prod_l(\phi_l)^{\mu_l}\over\Lambda^b}\right)^{2\over\Delta}
\,,
\ee
where $\mu_l$ is the (quadratic) Dynkin index of the representation $r_l$,
$\mu_G$ denotes the index of the adjoint representation, 
$\Delta=\sum_l\mu_l-\mu_G$ and $b=\half(3\mu_G-\sum_l\mu_l)$ is the coefficient
of the 1-loop $\beta$-function. In general the complete \nonpert
superpotential consists of a sum of terms of the form (\ref{Weff})
with different possible contractions of all gauge and flavor indices.
The relative coefficients of these terms cannot be fixed by symmetry
arguments but must be inferred from a different reasoning.

If the theory is confining at every point of the moduli space then the
superpotential must either vanish or be a smooth function of the confined 
degrees of freedom \cite{confine}. All such models with a smooth confining 
superpotential could be classified \cite{confine} as they have to verify the 
constraint $\Delta=2$. But only for some of these smoothly confining gauge 
theories containing tensor fields a dual description in terms of magnetic 
variables is known.

On the other hand many dualities for models including tensor fields have been
found once an appropriate tree-level superpotential for the tensors is added.
Let us review one example first studied by the authors of \cite{KuSch}. They
considered an $SU(N_c)$ gauge theory with $N_f$ quark flavors $Q$, $\bar Q$, an
additional matter field $X$ in the adjoint representation and a tree-level
superpotential $\Wtree=\Tr X^{k+1}$, where $k>1$ is some integer. This model
has a dual description in terms of an $SU(kN_f-N_c)$ gauge theory with $N_f$
quark flavors $q$, $\bar q$, an adjoint tensor $Y$, $kN_f^2$ singlets 
$M_{{\rm mag},j}$, $j=0,\ldots,k-1$ and tree-level superpotential
\be
\Wmag=\Tr Y^{k+1}\:+\:\sum_{j=0}^{k-1}M_{{\rm mag},k-1-j}qY^j\bar q\,.
\ee
For $N_c=kN_f-1$ the magnetic theory is completely higgsed and one expects the
electric theory to confine. Indeed, one finds that the 't Hooft anomaly
matching conditions are satisfied if the confined spectrum of the electric
theory is given by \cite{KSS}
\bea \label{confspec}
M_j &= &QX^j\bar Q\,,\quad j=0,\ldots,k-1\,,\\
B   &= &(Q)^{N_f}\cdots(X^{k-1}Q)^{N_f}(X^kQ)^{N_f-1}\,,\nonumber\\
\bar B &= &(\bar Q)^{N_f}\cdots(X^{k-1}\bar Q)^{N_f}(X^k\bar Q)^{N_f-1}\,,
\nonumber
\eea
where the gauge indices are contracted with a Kronecker delta for the mesons 
and with an epsilon tensor of rank $N_c$ for the baryons. In addition the
flavor indices of the baryons are contracted with an epsilon tensor of rank
$kN_f$ leaving $2N_f$ independent baryons. It is easy to see \cite{mklein} 
that these are exactly the degrees of freedom that are mapped under duality
on the magnetic singlets $\hat M_{{\rm mag},j}$, $\hat q$, $\hat{\bar q}$ that
stay massless after the Higgs effect. The matching of the 't Hooft anomalies
between the microscopic (i.e.\ quarks and gluons) and the macroscopic (i.e.\
confined) description of the electric theory can thus be seen as a consequence
of the anomaly matching between the electric and the magnetic theory. In this
sense confinement can be derived from duality.

\TABLE{
$\ba{|c|c|c|c|c|} \hline 
\rule{0mm}{3ex} &\multicolumn{4}{c|}{SU(N_c)}\\ \hline
\rule{0mm}{3ex} {\rm tensors} &adj &\Yasym+\Yasymb &\Ysym+\Ysymb 
                              &\Yasym+\Ysymb\\
\rule{0mm}{3ex} \Wtree &X^{k+1} &(X\bar X)^{k+1} &(X\bar X)^{k+1} 
                       &(X\bar X)^{2(k+1)}\\
\rule{0mm}{3ex} N_c &kN_f-1 &(2k+1)N_f-4k-1 &(2k+1)N_f+4k-1 &(4k+3)(N_f+4)-1\\
\rule{0mm}{3ex} \alpha &k &1 &2(k+1) &2(k+1) \\
\rule{0mm}{3ex} \beta_X &(k-1)N_c &k(N_f-1) &(k+1)(2k(N_f+2)-1) 
                &\msmall{2(k+1)((2k+1)(N_f+4)-3)}\\
\rule{0mm}{3ex} \beta_{\bar X} &  &k(N_f-1) &(k+1)(2k(N_f+2)-1) 
                &\msmall{2(k+1)((2k+1)(N_f+4)+1)}\\
\rule{0mm}{3ex} \gamma &-N_c &3-N_f &-(N_c+N_f+4) &-2(k+1)(N_f+4)\\ \hline
\ea$\phantom{xx}

$\ba{|c|c|c|c|c|} \hline \rule{0mm}{3ex} &\multicolumn{2}{c}{Sp(2N_c)} 
                        &\multicolumn{2}{|c|}{SO(N_c)}\\ \hline
\rule{0mm}{3ex} {\rm tensors} &\Ysym &\Yasym &\Yasym &\Ysym\\
\rule{0mm}{3ex} \Wtree &X^{2(k+1)} &X^{k+1} &X^{2(k+1)} &X^{k+1}\\
\rule{0mm}{3ex} N_c &(2k+1)N_f-2 &k(N_f-2) &(2k+1)N_f+3 &k(N_f+4)-1\\
\rule{0mm}{3ex} \alpha &2k+1 &1 &1 &k\\
\rule{0mm}{3ex} \beta &2k(N_c+1) &(k-1)(N_f-1) &2k(N_f+1)+4 &(k-1)(N_c-2k)\\
\rule{0mm}{3ex} \gamma &-(N_c+1) &3-N_f &1-N_f &-(N_c+2k)\\ \hline
\ea$
\caption{Gauge theories that confine in the presence of a
  tree-level superpotential. The microscopic spectrum consists of $N_f$
  quark flavors and additional fields transforming in tensor representations
  represented by their Young tableaux. The coefficients $\alpha$, $\beta$, 
  $\gamma$ refer to the powers in the \nonpert superpotential (\ref{Weff2}).}
}
\TABLE{
$\ba{|c|c|c|c|c|} \hline 
\rule{0mm}{3ex} &\multicolumn{4}{c|}{SU(N_c)}\\ \hline
\rule{0mm}{3ex} {\rm tensors} &2\,adj &adj+\Yasym+\Yasymb &adj+\Ysym+\Ysymb 
                              &adj+\Yasym+\Ysymb\\
\rule{0mm}{3ex} \Wtree &X^{k+1}+XY^2 &X^{k+1}+XY\bar Y &X^{k+1}+XY\bar Y 
                       &X^{k+1}+XY\bar Y\\
\rule{0mm}{3ex} N_c &3kN_f-1 &3kN_f-5 &3kN_f+3 &3k(N_f+4)-1\\
\rule{0mm}{3ex} \alpha &3k &k &4k &2k \\ \hline
\ea$\phantom{xx}

$\ba{|c|c|c|c|c|} \hline \rule{0mm}{3ex} &\multicolumn{2}{c}{Sp(2N_c)} 
                        &\multicolumn{2}{|c|}{SO(N_c)}\\ \hline
\rule{0mm}{3ex} {\rm tensors} &2\,\Yasym &\Yasym+\Ysym &2\,\Ysym 
                              &\Ysym+\Yasym\\
\rule{0mm}{3ex} \Wtree &X^{k+1}+XY^2 &X^{k+1}+XY^2 &X^{k+1}+XY^2 
                       &X^{k+1}+XY^2\\
\rule{0mm}{3ex} N_c &3kN_f-4k-2 &3kN_f-4k+2 &3kN_f+8k+3 &3kN_f+8k-5\\
\rule{0mm}{3ex} \alpha &(k) &(3k) &3k &k\\ \hline
\ea$
\caption{Gauge theories that confine in the presence of a
  tree-level superpotential. The microscopic spectrum consists of $N_f$
  quark flavors and additional fields transforming in tensor representations
  represented by their Young tableaux. The coefficient $\alpha$ refers to the 
  power in the \nonpert superpotential (\ref{Weff2}).}
}

It is straightforward to apply this idea to all gauge theory models of 
\cite{ILSdual,patterns} based on simple gauge groups. One first builds the 
gauge invariant composite operators whose expectation values span the moduli 
space, then finds the duality mapping between the electric and the magnetic
theory for these operators and finally applies this mapping to the completely
higgsed effective magnetic theory to obtain the confined spectrum of the
electric theory. In addition, in most cases at least some of the terms of the 
confining superpotential can be determined from the magnetic tree-level 
superpotential. To constrain the possible form of the confining superpotential
we would like to generalize the formula (\ref{Weff}) to the case of a \nonvan
tree-level superpotential. Therefore divide the matter fields into two subsets
$\{\phi_l\}=\{\bar\phi_{\bar l}\}\cup\{\hat\phi_{\hat l}\}$,
with $\{\bar\phi_{\bar l}\}\cap\{\hat\phi_{\hat l}\}=\emptyset$,
and add a tree-level term for the hatted fields:
\be \label{Wtreehat}
\Wtree=h\,\prod_{\hat l}\left(\hat\phi_{\hat l}\right)^{n_{\hat l}},
\ee
where $h$ is a dimensionful coupling parameter and the $n_{\hat l}$
are positive integers. To be invariant under all global symmetries the full
superpotential must be of the form \cite{mklein}
\be \label{Weff2}
W \propto\left({\prod_{\bar l}(\bar\phi_{\bar l})^{\mu_{\bar l}}\over\Lambda^b}
          \right)^\alpha\ \prod_{\hat l}\left(\hat\phi_{\hat l}
          \right)^{\beta_{\hat l}}\ h^\gamma\,,
\ee
where the powers $\alpha$, $\beta_{\hat l}$, $\gamma$ must verify the
following relations:
\bea \label{albega}
\gamma &= &1-\half\alpha\Delta\,,\nonumber \\
\beta_{\hat l} &= &\mu_{\hat l}\,\alpha+\gamma n_{\hat l}\,.
\eea

The calculation of the confining spectra and the confining superpotentials
for all simple group models of \cite{ILSdual,patterns} is performed in 
\cite{mklein}. The results are displayed in tables 1 and 2.

\section{A new confining model}
To illustrate the ideas presented above I would like to treat one of the models
of \cite{mklein} in more detail. Consider an $SU(N_c)$ gauge theory with 
$N_f+8$ quarks $Q$, $N_f$ antiquarks $\bar Q$, an \asym tensor $X$ and a 
conjugate symmetric tensor $\bar X$ and tree-level superpotential 
$\Wtree=h\,\Tr(X\bar X)^{2(k+1)}$. This is a chiral theory and the difference
between the number of quarks and antiquarks is required by anomaly freedom.
The gauge invariant composite operators are given by 
\bea \label{mesbarsas}
&& M_j=Q\bar Q_{(j)}\,,\quad
   P_r= Q\bar XQ_{(r)}\,,\ \bar P_r=\bar QX\bar Q_{(r)}\,,\nonumber\\
&&\quad\w Q_{(j)}=(X\bar X)^j Q\,,\ \bar Q_{(j)}=(\bar XX)^j\bar Q\,,
  \nonumber\\
&& \quad j=0,\ldots,2k+1\,,\quad r=0,\ldots,2k\,,\nonumber \\
&& \bar\cB^{(\bar n_0,\ldots,\bar n_{2k},n_0,\ldots,n_{2k+1})}
   =(\bar X(X\bar X)^kW_\alpha)^2\nonumber\\
&& \qquad \cdot(\bar XQ)^{\bar n_0}(\bar XQ_{(1)})^{\bar n_1}
           \cdots(\bar XQ_{(2k)})^{\bar n_{2k}}\nonumber\\
&& \qquad \cdot\bar Q^{n_0}\bar Q_{(1)}^{n_1}\cdots
           \bar Q_{(2k+1)}^{n_{2k+1}}\,,\nonumber\\
&& \quad\w \sum_{j=0}^{2k+1}n_j+\sum_{j=0}^{2k}\bar n_j=N_c-4\,,\\
&& B_n=X^n Q^{N_c-2n}\,,\quad 
   n=0,\ldots,\left\lceil{N_c\over2}\right\rceil\,,\nonumber\\
&& \bar B_{\bar n}=\bar X^{\bar n} \bar Q^{N_c-\bar n}
               \bar Q^{N_c-\bar n}\,,\quad \bar n=0,\ldots,N_c\,,\nonumber\\
&& T_i=\Tr(X\bar X)^i\,,\quad i=1,\ldots,2k+1\,,\nonumber
\eea
where the gauge indices are contracted with one epsilon tensor for the 
$\bar\cB^{(\cdots)}$, $B_n$ and with two epsilon tensors for the 
$\bar B_{\bar n}$.

The authors of \cite{ILSdual} found a dual description of this model in terms
of a magnetic $$SU((4k+3)(N_f+4)-N_c)$$ gauge theory, with $N_f+8$ quarks $q$, 
$N_f$ antiquarks $\bar q$, an \asym tensor $Y$, a conjugate symmetric tensor 
$\bar Y$ and singlets $M_{{\rm mag},j}$, $P_{{\rm mag},r}$, 
$\bar P_{{\rm mag},r}$ and tree-level superpotential 
$$\Wmag=-h\,\Tr(X\bar X)^{2(k+1)}+\ldots\,,$$ where the dots indicate terms
involving $M\mag$, $P\mag$, $\bar P\mag$. The duality mapping for the gauge
invariant operators is given by
$$M_j,\,P_r,\,\bar P_r\ \lra\ 
M_{{\rm mag},j},\,P_{{\rm mag},r},\,\bar P_{{\rm mag},r}\,,$$
\bea \label{barmapsas}
   \bar\cB^{(\bar n_i,n_j)} &\lra &\bar\cB\mag^{(\bar m_i,m_j)}\,,\nonumber\\
   \w && m_j=N_f-n_{2k+1-j}\,,\nonumber\\
      &&\bar m_j=N_f+8-\bar n_{2k-j}\,,\\ 
   B_n &\lra &B_{{\rm mag},m}\,,\nonumber\\\
   \w && m=(2k+1)(N_f+4)-2-n\,,\nonumber\\ 
   \bar B_{\bar n} &\lra &\bar B_{{\rm mag},\bar m}\,,\nonumber\\
   \w && \bar m=2(2k+1)(N_f+4)+4-\bar n\,.\nonumber
\eea
The dependence on the mass scale $\mu$ has been suppressed.

For $N_c=(4k+3)(N_f+4)-1$ the magnetic theory is completely higgsed and the 
electric theory confines with low-energy spectrum given by the composite 
fields 
\bea \label{confspecsas} 
&&M_j\,,\ P_r\,,\  \bar P_r\,,\nonumber\\
&&\quad j=0,\ldots,2k+1\,,\quad r=0,\ldots,2k\,,\nonumber\\
&&B\equiv B_{(2k+1)(N_f+4)-2}\,,\nonumber\\ 
&&\bar B\equiv\bar\cB^{(N_f+8,\ldots,N_f+8,N_f,\ldots,N_f,N_f-1)}\,,\\ 
&&\bar b\equiv \bar B_{2(2k+1)(N_f+4)+3}\,,\nonumber
\eea
of eqs.\ (\ref{mesbarsas}). From (\ref{barmapsas}) one finds the mappings 
$B,\bar B\lra q,\bar q$ and $\bar b\lra \bar Y$. One color component of 
each of the fields $q$, $\bar q$, $\bar Y$ together with the meson singlets 
are exactly the degrees of freedom that stay massless after breaking the 
magnetic gauge group.

As a further consistency check let us consider deformations of the theory
along the flat directions corresponding to large expectation values of the 
baryons $B$, $\bar b$. A large VEV of $B$ breaks the gauge symmetry to 
$Sp(2((2k+1)(N_f+4)-2))$ \cite{ILSdual}. The low-energy theory contains 
$2(N_f+4)$ quarks $Q$, a symmetric tensor $X$ and tree-level superpotential 
$\Tr X^{2(k+1)}$. This model is known to show confinement \cite{newconf}. 
A large VEV of $\bar b$ breaks the gauge symmetry to $SO(2(2k+1)(N_f+4)+3)$ 
\cite{ILSdual}. The low-energy theory contains $2(N_f+4)$ quarks $Q$, an \asym
tensor $X$ and tree-level superpotential $\Tr X^{2(k+1)}$. This model is known
 to show confinement \cite{newconf}.

The effective low-energy superpotential of the magnetic theory contains the 
terms $M_{2k+1}q\bar q\ +\ P_{2k}q\bar Yq$. 
We thus expect that the confining superpotential of the electric theory has 
terms proportional to $\bar B M_{2k+1} B$, $BBP_{2k}\bar b$. The detailed
analysis gives \cite{mklein}
\bea \label{Wconfsas}
    W &= &{\bar B M_{2k+1} B\over h^{2(k+1)(N_f+4)}\,
       \Lambda^{2(k+1)((8k+5)(N_f+4)-2)}}\nonumber\\
      &&\ +\ {BBP_{2k}\bar b\over h^{2(N_f+4)-1}\Lambda^{2((8k+5)(N_f+4)-2)}}\\
      &&\ +\ \ldots\,,\nonumber
\eea
where the dots stand for possible further terms that could be generated by
instanton effects in the completely broken magnetic gauge group.

\section{Conclusion}
I have shown how the \nonab duality of $N=1$ supersymmetric gauge theories 
discovered by Seiberg can be used to find new models that confine in the 
presence of an appropriate superpotential.
This is a very interesting application of the proposed duality because
it enables us to obtain \nonpert results for the electric theory by
a perturbative calculation in its magnetic dual. Confinement in the
electric theory can be understood from the Higgs phase of the magnetic
theory. The confining spectrum can easily be derived from the
duality mappings of gauge invariant operators. For $SU$ and $SO$ 
gauge groups one also obtains the form of the confining superpotential
by applying these mappings to the magnetic tree-level superpotential.
To determine the full confining superpotential one needs to include
instanton corrections in the completely broken magnetic gauge group.
For $Sp$ gauge groups the tree-level superpotential of the completely 
higgsed magnetic theory vanishes. In this case the whole magnetic 
superpotential is \nonpert and therefore more difficult to obtain.

\acknowledgments
I would like to thank the organizers for this very interesting workshop
and for their warm hospitality. I would also like to thank  L.~Ib\'a\~nez, 
\linebreak
C.~Cs\'aki and H.~Murayama for helpful discussions and e-mail correspondence. 
This work is supported by the European Union under grant FMRX-CT96-0090.

\end{document}